# Explaining the Color Distributions of Globular Cluster Systems in Elliptical Galaxies

Suk-Jin Yoon[1,2,3], Sukyoung Ken Yi[1,2] & Young-Wook Lee[1]

[1]Department of Astronomy & Center for Space Astrophysics, Yonsei University, Seoul 120-749, Korea

[2]Astrophysics, University of Oxford, Keble Road, Oxford OX1 3RH, UK

[3]To whom correspondence should be addressed. E-mail: sjyoon@galaxy.yonsei.ac.kr

**The colors of globular clusters in most of large elliptical galaxies are bimodal. This is generally taken as evidence for the presence of two cluster subpopulations that have different geneses. However, here we find that, because of the non-linear nature of the metallicity-to-color transformation, a coeval group of old clusters with a unimodal metallicity spread can exhibit color bimodality. The models of cluster colors indicate that the horizontal-branch stars are the main drivers behind the empirical non-linearity. We show that the scenario gives simple and cohesive explanations for all the key observations, and could simplify theories of elliptical galaxy formation.**

One of the most outstanding discoveries from observations of elliptical galaxies over the last decade is the bimodal color distribution of globular clusters − gravitationally bound collections of millions of stars (*1-8*). The phenomenon is widely interpreted as evidence of two cluster sub-systems with distinct geneses within individual galaxies (*9*). However, given many ways of forming clusters in elliptical galaxies, it is quite surprising that the cluster color distributions behave in an orderly way. For instance, the numbers of blue and red clusters in large galaxies are roughly comparable (*1-8*); blue and red clusters are old (> 10 billion years) and coeval (*9*), and differ systematically in spatial distribution and kinematics (*5,10-16*); and their relative fractions and peak colors strongly correlate with host galaxy properties (*2-8*). Here, we propose a simpler solution that does not necessarily invoke distinct cluster sub-systems and has a sound basis both on the empirical and theoretical relations between metallicity and colors.

A recent observation (*8*) reveals that the $g−z$ color (*17*) of clusters correlates with their [Fe/H] (*18*) (Fig 1A). The observed relation is tight enough to show a significant departure from linearity with a slope rapidly changing at [Fe/H] ≈ −1.0. A closer inspection suggests that they might follow an inverted S-shaped "wavy" curve with a quasi-inflection point at [Fe/H] ≈ −0.8. To examine this indication, we overplotted predicted colors of 13-Gyr clusters from two different models (*19,20*). While there is good agreement between the models, they predict systematically redder colors for the [Fe/H] ≈ –1.0 clusters which serve as key part of the possible invert S-shape of the observed relation. We present our equivalent 13-Gyr models with the [Fe/H] grid spacing of Δ[Fe/H] = 0.1 (Fig. 1B), a resolution sufficient to sample the region of [Fe/H] ≈ –1.0. The main asset of our model is the consideration of the systematic variation of the mean color of horizontal-branch (HB) stars as a function of [Fe/H] (*21-24*). The version of our model that excludes the prescription for the systematic HB variation, although showing agreement with the other models, does not match the observations as well. However, by including the realistic HB variation we can reproduce the clusters with [Fe/H] ≈ –1.0 and, in turn, the observed wavy feature. This suggests that the wavy feature along the sequence defined by the observed clusters is real.

The observed wavy feature in the metallicity-color relation is a result of two complementary effects (Fig. 1B): (a) the integrated color of the stars prior to the HB stage (i.e., the main-sequence and red-giant-branch) is a non-linear function of metallicity at given ages, showing a mild departure from linearity at lower metallicity, and (b) the color of the HB changes at a brisk pace between [Fe/H] ≈ –0.6 and –0.9, further strengthening the departure from linearity. As a result, the color becomes several times more sensitive to metallicity between [Fe/H] = –0.6 and –0.9, resulting in a quasi-inflection point at [Fe/H] ≈ –0.8. The former effect is visible in all models, but the latter effect is present only in our model (Fig. 1A, B). We illustrate the well-known effect of metallicity on the systematic HB color variation (Fig. 1C). The color of HB varies abruptly between [Fe/H] = –0.6 and –0.9 where the HB just departs from the red-clump position. The physics of this phenomenon is described in (*24*). The Galactic globular clusters with HB morphology similar to the synthetic models include NGC 6624, NGC 104, NGC 6638, and NGC 5904 for [Fe/H] = –0.4, –0.7, –1.0, and –1.3, respectively (*21*). We note that, since the range of the rapid change is as small as ~0.3 in [Fe/H], the wavy feature would not be present in the models with a [Fe/H] grid spacing larger than 0.3.

This non-linear nature of the empirical relation between intrinsic metallicity and its proxy, colors, may hold the key to understanding the color bimodality phenomenon: the wavy feature brings about the bimodality by projecting equi-distant metallicity intervals near the quasi-inflection point onto larger color intervals. To scrutinize this "projection effect", we have performed a Monte Carlo simulation using 100,000 coeval clusters (Fig. 2). For [Fe/H] distributions, we make a simple assumption of a broad Gaussian with standard deviation $\sigma_{[Fe/H]}$ = 0.5 dex (Fig. 2A, E). For a direct comparison with typical galaxies NGC 4649 and M87, we adopt the mean [Fe/H] of their entire globular clusters, $\langle[Fe/H]\rangle_{GC}$ ($\approx -0.65$), as inferred from the observed mean $g-z$ color ($\approx 1.3$) (*8*). Next, the $g-z$ color of each cluster is obtained using our theoretical [Fe/H]-$g-z$ relation for the 13-Gyr population. In the color-magnitude diagrams of 1,000 randomly selected clusters (Fig 2B, F), two vertical bands of clusters are immediately visible at $g-z \approx 0.9$ and 1.4. The resultant color histograms of 100,000 clusters (Fig. 2C, G) clearly show prominent dips near their centers, reproducing the observed histograms of NGC 4649 and M87 (Fig. 2D, H). For comparison, the color distribution (Fig. 2G) that is obtained using a simple linear (straight line) fit to the [Fe/H] vs. $g-z$ data is in clear conflict with the observation. It is interesting that the quasi-inflection points are also apparent in other combinations of bandpasses such as *V–I* (Fig. 2I). As a result, using the identical [Fe/H] distribution to that in Fig. 2E, the model (Fig. 2J) reproduces the observed *V–I* histogram (*5*) (Fig. 2L). These results are in good agreement with detailed studies indicating that the mean ages of both blue and red clusters in individual galaxies are old (> 10 Gyr) and comparable within a couple of Gyr (*9*). Moreover, the comparable numbers of blue and red clusters found in large galaxies are obviously due to the fact that the dip colors of the histograms corresponds to the midpoint of the color spanned by clusters in large galaxies.

Although photometry is used commonly to infer the metallicities of clusters, it is no substitute for spectroscopy. We wonder whether the observed distribution of metal indices such as *Mg b* absorption line (near 5170 Å) can also be explained by the projection effect. The simulation targets the M87 cluster system, which has a clear color bimodality (Fig. 2H, L) (*8,11*) and the largest cluster sample with measured metal indices (*25*). The observed *Mg b* distribution is, although not bimodal, highly asymmetric (Fig. 2P), which is often viewed as the sum of two cluster sub-systems. Since the absorption indices trace more directly the element abundance than colors do, the model predicts a relatively weaker wavy feature along the [Fe/H]-*Mg b* relation (Fig. 2M), which is in good agreement with the observation (*26*). When the identical [Fe/H] distribution to that in Fig. 2E, I is used, the model (Fig. 2O) reproduces successfully the observed *Mg b* histogram that has a broad metal-poor peak with a metal-rich tail (Fig. 2P). For comparison, the color distribution (Fig. 2O) that is obtained using a simple linear fit to the [Fe/H] vs. *Mg b* data is in clear conflict with the observation. We therefore conclude that the projection effect is also at work in the *Mg b* distribution of M87 clusters.

There is a growing body of evidence that the color distributions of globular cluster systems are closely linked to the host galaxy luminosity (*2-8*). The number fraction of red clusters and the mean colors of both blue and red clusters increase progressively for more luminous galaxies. We have found that the projection effect can explain these intriguing trends as well. To simulate the color distributions as a function of host galaxy $B$-band luminosity ($M_B$), we have adopted $\langle$[Fe/H]$\rangle_{GC}$ for each cluster system of various $M_B$ on the basis of the empirical $\langle$[Fe/H]$\rangle_{GC}$-$M_B$ relation (*8*) (Fig. 3D). The resulting color histograms (Fig. 3B) show that, in proceeding to lower luminosities, the distributions are more consistent with being predominantly blue-peaked. This is in good accordance with the observations (*8*) (Fig. 3C).

We compare the observed quantities across galaxy luminosity with those obtained from the model histograms. The observed link of the red cluster fraction (Fig. 3E) and the blue and red peak colors (Fig. 3F) to the host galaxy luminosity are well reproduced by means of the projection effect. Interestingly, the zero-point for red cluster colors is on average ~0.1 magnitude redder than the observations. This can be explained (Fig. 3A, F) if red clusters are slightly younger than blue ones (by ~2 Gyr) within the current uncertainty in cluster age dating (*9*), or if red clusters with [Fe/H] > −0.5 have an extended blue HB component as observed in the Galactic counterparts (*27,28*). Besides, the observed slope for red clusters is known to be 4.6 times steeper than for blue clusters (*8*) (Fig. 3F). This is probably because the slope in the [Fe/H]-$g-z$ relation is steeper by such a factor in the metal-rich regime (Fig. 3A). This effect also naturally explains the observed larger color spread of red clusters at given $M_B$ and the larger color ranges of clusters in brighter galaxies. We therefore conclude that the projection effect and the variation of [Fe/H] distribution are the main drivers behind the properties of the cluster color distribution as a function of host galaxy luminosity.

It is now well established that blue and red clusters show differences in spatial distribution and kinematics within individual galaxies (*5,10-16*). Red clusters appear to be more centrally concentrated and of lower velocity dispersion. The phenomenon can be considered as a close analogy with the aforementioned link between the color distributions and the host galaxy luminosities. Observational evidence (*11,25*) indicates higher $\langle$[Fe/H]$\rangle_{GC}$ toward the galaxy center. If this trend is employed in the simulation, then the projection effect causes naturally the red clusters to be more popular toward the galaxy center. Besides, kinematic studies indicate that the systematically lower velocity dispersion of red clusters is simply a consequence of the red cluster number density that is higher in the galaxy center (*16*). Thus, it appears that the differences between blue and red clusters in spatial distribution and kinematics are fully consistent with the projection effect explanation.

It has been a popular view that the presence of two discrete cluster sub-systems within individual galaxies is responsible for the color bimodality. Our own Galaxy, which possesses two globular cluster sub-systems with different metallicity and kinematics (*29*), has served as a typical example of the case. Whether elliptical galaxies and the Milky Way have an equivalent cluster formation history is an outstanding issue, however. One may argue that the Milky Way is not an archetypal host of the color bimodality found in elliptical galaxies, because the origin of the metal-rich component in the Galactic globular cluster system is apparently more complicated (*29-31*). With true metallicity being bimodal, color bimodality would be strengthened further. But the essence of our explanation is that we do not need to invoke two distinct metallicity groups to explain the observed level of color bimodality in elliptical galaxies. Further spectroscopic observations are definitely needed to obtain true metallicity distributions of globular cluster systems in elliptical galaxies from high signal-to-noise spectra.

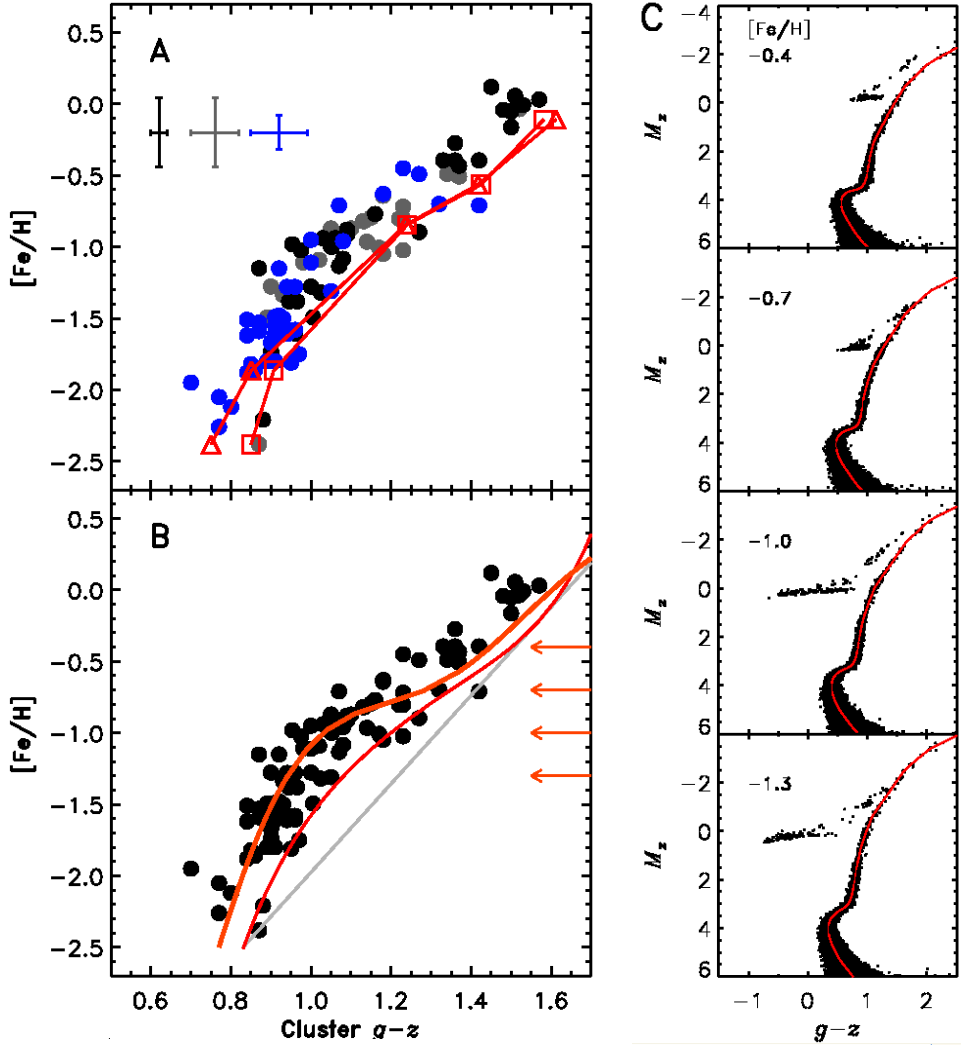

**Fig 1.** Correlation between iron abundance [Fe/H] and *g–z* for globular clusters in our Galaxy and two large elliptical galaxies, M49 and M87. (**A**) The 40 low-extinction ($E_{B-V}$ < 0.3) Galactic clusters (blue symbols), 33 M49 and M87 clusters with ACS (the Advanced Camera for Surveys) Virgo Cluster Survey photometry (black symbols), and 22 M49 and M87 clusters with SDSS (Sloan Digital Sky Survey) photometry (grey symbols) are obtained from (*8*). The typical errors have been estimated from references in (*8*) and shown with the same color code. In order to take into account the non-solar α-element (O, Mg, Si, S, Ca, and Ti) abundance of clusters, the α-element with respect to iron, [α/Fe] = 0.3 models in (*32*) are used to amend [Fe/H]. Simple stellar population models for 13-Gyr clusters are overlaid. Red squares and triangles represent the predictions from (*19*) and (*20*), respectively. (**B**) The identical observed data in (A) are shown. Our equivalent 13-Gyr population model (thick orange line) with the finer grid spacing (Δ[Fe/H] = 0.1) is overlaid. The thin red line is for the model without inclusion of the HB variation. The grey line is the linear connection between *g–z* = 0.83 and 1.70 which assists in estimating the degree of the departure from the linearity. Arrows denote [Fe/H]'s for which the color-magnitude diagrams are shown in (C). (**C**) Synthetic color-magnitude diagrams for 13-Gyr clusters with various [Fe/H]. Red loci are the model isochrones.

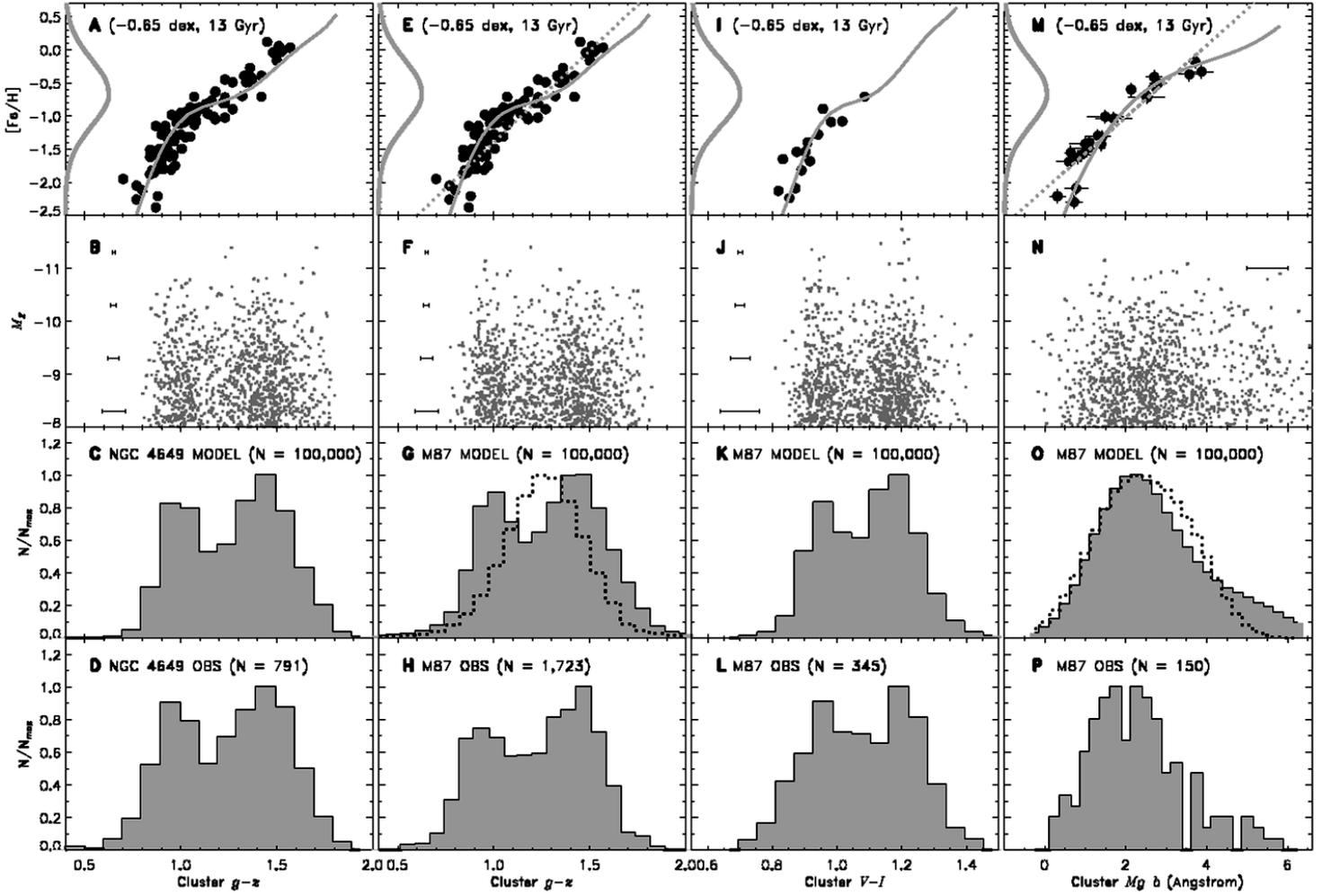

**Fig 2.** Monte Carlo simulations of globular cluster color distribution. (**A**) Same as Fig. 1B. The metallicity distribution of $10^5$ model clusters is shown along the *y*-axis (thick grey line) and their mean metallicity and age are denoted in parenthesis. (**B**) Color-magnitude diagram of 1,000 randomly-selected model clusters of 13 Gyr. Cluster *g–z* is transformed from [Fe/H] using the theoretical relation shown in (A). For the integrated *z*-band absolute magnitude, $M_z$, a Gaussian luminosity distribution ($\langle M_z \rangle = -8.19$ and $\sigma_z = 1.03$) is assumed (*7*). The observational uncertainty in *g–z* as a function of $M_z$ is taken into account based on the observations (*5*). (**C**) The color histogram of $10^5$ model clusters of 13 Gyr. (**D**) The observed color histogram for 791 clusters in NGC 4649 (*8*). (**E** to **H**) Same as (A) to (D), but for 1723 clusters in M87 (*8*). The dotted histogram in (G) represents the distribution that is obtained using a simple linear fit shown by dotted line in (E). (**I** to **L**) Same as (E) to (H), but for the *V* (~5550 Å)–*I* (~8140Å) color. The data in (I) are from (*33*). The observed histogram is for 345 clusters in M87 (*5*). (**M** to **P**) Same as (E) to (H), but for *Mg b*. The data in (M) are from (*26*). The dotted histogram in (O) represents the distribution that is obtained using a simple linear fit shown by dotted line in (M). The observed histogram is for 150 clusters in M87 (*25*). The uncertainty of 0.5 Å is assumed (*25*).

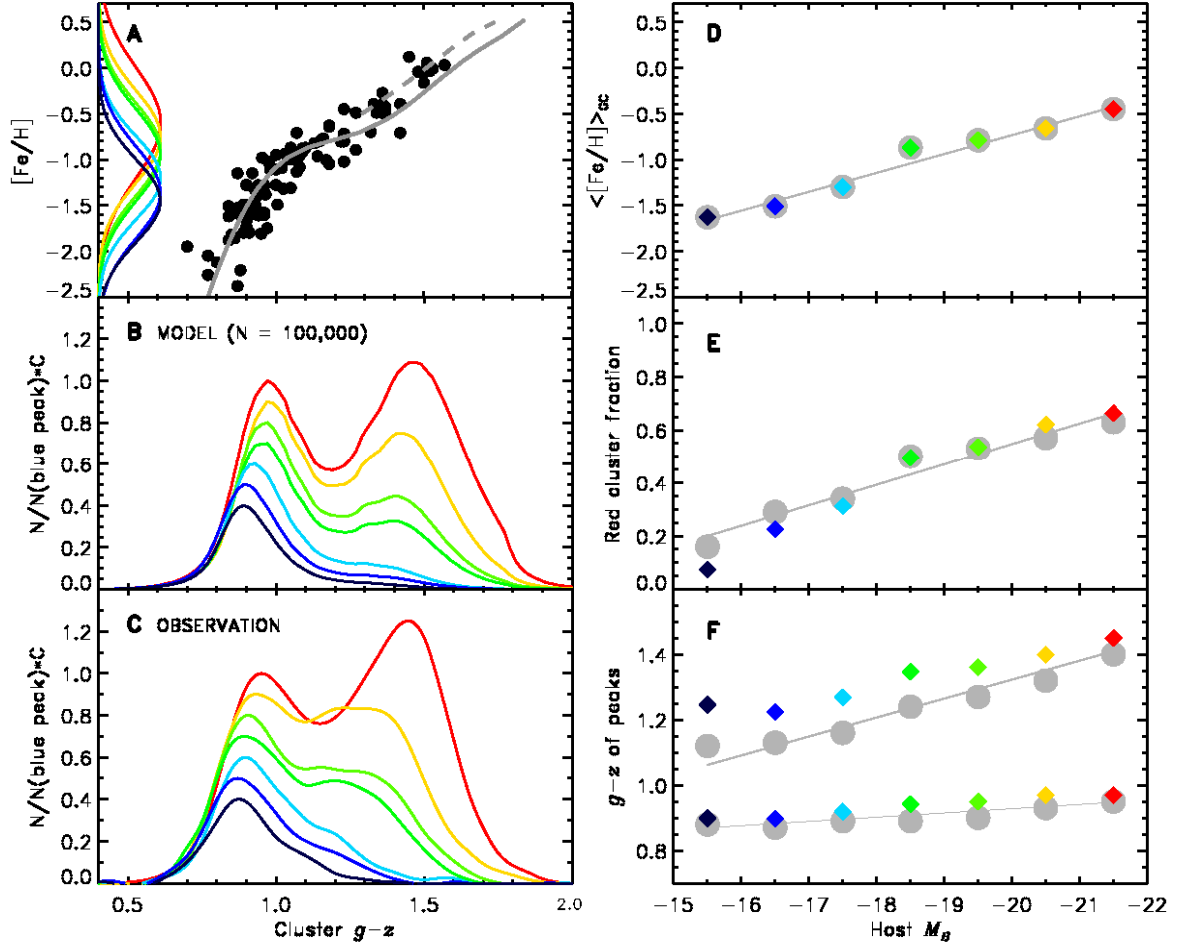

**Fig 3.** Monte Carlo simulations of globular cluster color distributions for various host galaxy luminosity. (**A**) Same as Fig. 1B. Various metallicity distributions of $10^5$ model clusters are shown along the *y*-axis. The values of $\langle$[Fe/H]$\rangle_{GC}$ are adopted from the observations shown in (D). Both the 11-Gyr model and the model with extended blue HB component (30 % in number) reproduce better the observed red clusters with *g*–*z* > 1.3 (dashed line). (**B**) The model color histograms of cluster systems for seven bins of host galaxy magnitude. The histograms are normalized by the number of blue clusters and multiplied by constants, *C*, for clarity. The magnitude bins are 1 magnitude wide and extend from $M_B$ = –21 (red, *C* = 1.0) to –15 (purple, *C* = 0.4). (**C**) The observed color histograms of clusters in the same magnitude bins, multiplied by corresponding constants, *C*. (**D** to **F**) The observational data – the mean metallicity $\langle$[Fe/H]$\rangle_{GC}$, the fraction of red clusters, and the peak colors of blue and red clusters – as functions of $M_B$ (*8*) are denoted by large grey circles. The solid line in each panel is the least-square fit. The simulation results are marked by small diamonds with the same color code as in (A) to (C).

9